# A Better Understanding of Multi-User Cooperation: A Tradeoff between Transmission Reliability and Rate

*A short review for "Capacity-Outage-Tradeoff for Cooperative Networks"*
Edited by W. Guo, and S. Wang, and X. Chu

Cooperative communications involves nodes that exchange information and transmitting jointly to a common destination [1]. Whilst different users sharing content is increasingly common on the network-and data-layer, it has so far not occurred on the physical wireless interface. Cooperative transmission has the underlying potential to dynamically tradeoff data rate with reliability, depending on the multi-media content transmitted. The key prohibitor of this technology is a lack of understanding of the penalties and benefits in sharing signals on a physical level.

The rationale for cooperative transmission is that by transmitting the same (or similar) data along different channels, the stochastic nature of multipath fading can be exploited. This has proven to be especially effective in quasi-static (slow) fading channels, where the information coding-length is smaller the fading variation period. In such a channel, the achievable capacity at arbitrarily high reliability is zero, and reducing the outage probability is a challenge [2].

A key drawback with repetitively transmitting the same information along multiple channels is that the resource utilization of the channel is inefficient [2]. Therefore, given a fixed power and spectrum constraint, increased cooperation leads to a decreased amount of power and bandwidth per transmission in the cooperation process.

The aim of the paper [3] is to present a tradeoff between data throughput and transmission reliability for cooperative transmission. This tradeoff can assist to achieving content aware cooperation on the physical layer, whereby depending on the transmission reliability requirements of the multi-media content; maximum data rate can be achieved by selecting the optimal number of cooperation partners.

In the reviewed paper [3], the authors first propose the novel relationship between achieving greater transmission reliability through cooperation and the associated reduction in transmission efficiency. The primary contribution of the paper is formalizing a tradeoff between transmission reliability (outage probability) and throughput (capacity). A key distinction between this work and existing literature is that this paper's signal transmission employs realistic modulation and forward-error-correction (FEC) codes. This offers a realistic insight compared to the commonly used Shannon expression (infinite code length), which has been shown to be over-optimistic and can lead to misleading results [4].

The analysis performed in the paper uses theoretical expressions based on the bit-error-rate of transmitted information, which is reinforced by Monte-Carlo numerical simulation results. The specific cooperation protocol considered is known as Decode-and-Forward (DF), and its key advantages are: no noise amplification and no channel estimation at the relays.

The key conclusion from the tradeoff is that increased cooperation doesn't monotonically lead to increased transmission reliability. The relationship is in fact convex, and for any given system setup and user (channel conditions and transmission scheme), there exists an optimal set of cooperation partners which maximizes the transmission reliability. Furthermore, maximizing the reliability doesn't lead to maximizing the throughput. Therefore, the system designer or the user, need to tradeoff:

- Throughput,
- Reliability,

depending on the higher-layer multi-media content transmitted. For example, speech may require high transmission reliability, but a very low throughput rate. Therefore, for a user with a poor quality channel, cooperating with a large number of partners is desirable. On the other hand, for downloading data, the throughput rate is more important than reliability.

The second contribution of the paper is optimizing the system-level outage-capacity performance through partner selection schemes, which draws on the authors' previous work [5]. The authors found theoretical expressions on partner selection, based on the channel conditions and the desirable performance metrics. For a topology, where all nodes are roughly equal-distant to each other (symmetrical), it was found that the optimal number of partners is directly proportional to both the mutual channel strength, and the transmission scheme's Signal-to-Noise Ratio (SNR) threshold [3]. For a topology, where all nodes arbitrarily located (asymmetrical), it was found that the optimal number of partners can be found using a step-by-step numerical solution [3].

The third contribution of the paper is that given the selected partners, power allocation can be performed, so that power is optimally distributed amongst the cooperative transmission slots, which can maximize the transmission reliability and data rate. The results show that this can actually lead to requiring fewer cooperation partners, which implies that a joint optimality between partner selection and power allocation can be achieved. Future work can focus on joint optimality solutions, as well as how to combine media streams of different requirements into the same multi-user cooperation cycle.

In summary, the paper has presented a tradeoff between data throughput and transmission reliability for multi-user cooperative protocols. This tradeoff can achieve content aware cooperation on the physical layer, whereby depending on the transmission reliability requirements of the multi-media content; maximum data rate can be achieved by selecting the optimal number of partners. There remains significant work to be done on joint partner selection and power allocation strategies, as well as how to combine different multi-media contents into the same cooperation cycle.

**Acknowledgement:**

The R-Letter Editorial Board thanks Xiaoli Chu for nominating this work.